\documentclass[sigconf, 10pt, nonacm]{acmart}
\usepackage{subcaption}
\usepackage{listings}

\renewcommand\footnotetextcopyrightpermission[1]{} 

\begin{document}
\title{Towards application-specific query processing systems}

\author{Dimitrios Vasilas}
\affiliation{%
  \institution{Scality}
  \institution{Sorbonne Universit\'e - LIP6 \& Inria}
}
\email{dimitrios.vasilas@lip6.fr}

\author{Marc Shapiro}
\affiliation{%
  \institution{Sorbonne Universit\'e - LIP6 \& Inria}
}
\email{marc.shapiro@acm.org}

\author{Bradley King}
\affiliation{%
  \institution{Scality}
}
\email{brad.king@scality.com}

\author{Sara S. Hamouda}
\affiliation{%
  \institution{Sorbonne Universit\'e - LIP6 \& Inria}
}
\email{sara.hamouda@inria.fr}

\begin{abstract}
Database systems use query processing sub-systems for enabling efficient query-based data retrieval.
An essential aspect of designing any query-intensive application is tuning the query system to fit the application's
requirements and workload characteristics.
However, the configuration parameters provided by traditional database systems do not cover the design decisions
and trade-offs that arise from the geo-distribution of users and data.
In this paper, we present a vision towards a new type of query system architecture that addresses this challenge by
enabling query systems to be designed and deployed in a per use case basis.
We propose a distributed abstraction called Query Processing Unit that encapsulates primitive query processing tasks,
and show how it can be used as a building block for assembling query systems.
Using this approach, application architects can construct query systems specialized to their use cases, by
controlling the query system's architecture and the placement of its state.
We demonstrate the expressiveness of this approach by applying it to the design of a query system that can flexibly
place its state in the data center or at the edge, and show that state placement decisions affect the trade-off between
query response time and query result freshness.
\end{abstract}

\maketitle

\section{Introduction}
\label{sec:intro}
A major requirement of user-facing services is providing fast responses to user requests \cite{TAO2013, Noria2018}.
Serving user requests involves communication from a user's device to an application server,
and then potentially querying data from a backend store.

It is well known that achieving good query processing performance requires tunning the query processing system to the
needs of different use cases.
To achieve that, query systems expose configuration parameters and knobs such as selecting which indexes to materialize
and choosing between consistent and lazy index maintenance.
In addition, query systems are able to generate query execution plans adapted to each individual query
(query optimization).

However, the users and data of most web services nowadays are geographically distributed across the globe.
In the context of geo-distribution, the objectives of fast query response, consistent results, and low operational
cost are inherently conflicting and create trade-offs.

\textbf{Query response time.} In a query processing system in which data are distributed across multiple data centers,
query execution may involve communication round-trips across data centers, incurring overhead to query response time.

\textbf{Query result consistency.} Query systems aim at returning up-to-date results, which requires keeping their derived
state (e.g. indexes, materialized views) in sync with the base data.
In the context of geo-distribution this is often not practical due to the resulting overhead to write operations.
The alternative of updating derived state asynchronously can lead to returning stale query results.

\textbf{Operational cost.} Network resources across geographically distributed sites are often limited and costly,
therefore the amount of data transferred by the query system across sites affects network consumption and hence the
system's operational cost.

While techniques for reducing query response time have been studied extensively and applied in commercial database systems,
the aspects of query result consistency and operational cost have not been sufficiently explored.

A crucial design choice that affects the above trade-offs is the placement of the query processing system's state and computations.
Existing database systems provide little flexibility for configuring query processing state and computation
placement \cite{CockroachDB:topology}.

We argue that, achieving the right balance between low response time, up-to-date query results, and low operational cost,
requires additional tunning mechanisms.
Moreover, this balance is different across geo-distributed application due to their diverse characteristics and
requirements.

To address this need, we propose an architectural design pattern that enables application designers to navigate the
design space of geo-distributed query processing by giving them control over the placement of the query processing
system's state and computations.
The key idea is that query processing can be decomposed into basic tasks that can be encapsulated by independent components.
We present an architecture component abstraction, called Query Processing Unit (QPU), that embodies this principle.
The QPU abstraction defines a set of interfaces and a communication protocol.
Different instantiations of the abstraction implement different functionalities while conforming to the common specification.
Query processing systems are constructed by interconnecting QPU instances in a modular microservice-like architecture.

We demonstrate the expressiveness of the proposed approach by applying it to the design of a middleware system that
maintains materialized views in order to speed up query processing.
We show how the proposed approach enables materialized views to be flexibly placed in the cloud or at the edge according
to the application's access patterns and requirements.

This work includes the following contributions:
\begin{itemize}
  \item We propose an architectural design pattern for constructing and deploying geo-distributed query processing
  systems tailored to the characteristics and requirements of individual applications.
  \item We realize this design pattern by introducing an abstraction termed Query Processing Unit aimed to enable
  the design of modular query processing systems.
  We show how a geo-distributed query system can be constructed using the QPU abstraction (Section \ref{sec:qpu}).
  \item We demonstrate the expressiveness of the QPU-based architectural design and the flexibility that can be
  obtained from its use by applying it to the design of an application that maintains materialized views at the edge
  (Section \ref{sec:mv_edge}).
\end{itemize}

\section{Background}
\label{sec:background}
Internet services typically rely on a two-tiered backend architecture to serve user requests.
Data is stored persistently in a database (data storage tier).
A query processing tier is used on top of the database to provide the required query processing capabilities to the
service.

The query processing tier can serve two roles.
In cases where the database does not support querying, the query processing tier is responsible for providing querying
capabilities to the service, for example by building secondary indexes.
Alternatively, the query processing tier's responsibility may be to improve query scalability and performance by
maintaining pre-computed or cached query results.
Typically, the query processing layer maintains derived state (indexes, caches, materialized views) for serving queries.
Changes to the base data need to be propagated and applied to the derived state (state maintenance).
State maintenance is often asynchronous, and therefore derived state is eventually consistent relative to the base data.

We model the system as a collection of \textit{sites}.
We define a site as a group of servers located in the same geographic location.
Network communication latency among servers within a site is significantly lower than latency among servers
on different sites (a few milliseconds compared to tens to hundreds of milliseconds).
Moreover, network resources within a site cost significantly less compared to network resources across sites.
This is in accordance with cloud service pricing models where cross-site data transfer is more expensive
than data transfer within a site. \cite{ec2:pricing}.
A site can correspond either to a data center, or to a collection of edge nodes that form a tier along with the user
devices located geographically close to them.

In this paper, we examine the challenges and design decisions involved in the design of the query processing tier.
We assume that the query processing tier is deployed on top of an already existing data storage tier architecture.

\section{Design}
\label{sec:qpu}
In this section we present our main contribution, an architectural design pattern for constructing and deploying
geo-distributed query processing systems.
This pattern decouples the query system from the storage architecture, and shifts the task of designing and
implementing it from the database designer to the application architect.

In particular, our design is based on the following objectives:
\vspace{-2.5mm}
\begin{itemize}
  \item \textbf{Declaratively defined architecture.} The query system architecture should not be predefined.
  Rather, system architects should be able to construct query processing architectures in a per use case basis, and
  have control over the query processing techniques that the system employs.
  The query system design process should include decisions such as selecting the query processing state
  partitioning and replication schemes, and whether to use caching.
  In addition, the architecture should not make assumptions about the distribution scheme of the base data.
  \item \textbf{Flexible component placement.} The system designer should have fine-grained control over the placement
  of the query system's state and computations.
\end{itemize}

\subsection{The Query Processing Unit abstraction}
\label{subsec:qpu}
We enable these objectives using an assembly-based design strategy.
The key idea is that query processing systems can be constructed by assembling composable building blocks that
encapsulate primitive query processing tasks.
In that way, complex query processing tasks can be expressed through composition of simple building blocks.
To enable this architecture design pattern, we introduce an architecture component abstraction, termed Query Processing
Unit (QPU).

The QPU abstraction has the role of an architecture component template: it defines a set of properties including
interfaces, functionalities, and communication patterns.
Different instantiations (classes) of the QPU abstraction can be defined and implemented, but all need to conform
to the properties defined by the abstraction.
Instances of these QPU classes are the concrete building blocks that can be composed to construct query processing
systems.
In the rest of this section we use the terms \textit{query processing unit}, \textit{QPU}, and \textit{unit}
interchangeably.

We define the query processing unit as a long-running process with the following properties:
\begin{itemize}
  \item \textbf{Query processing state (optional).} A QPU can either maintain state, or be stateless.
  Stateful QPUs can express query processing tasks such as indexing (in which case the state is an indexing data
  structure), caching, and aggregation computations.
  Stateless QPUs can express tasks such as filtering and projections.
  \item \textbf{Query API.} Each QPU exposes an API for receiving and serving queries.
  When this API is called, a stream connection is established between the QPU and the caller.
  Query result entries are sent though the stream as stream records.

  The unit implements a \textit{query processing computation} that is invoked when the query API is called.
  The implementation of this computation is different for each QPU class.
  However, any implementation can use two functionalities: reading from the unit's state and performing downstream
  queries.
  \item \textbf{Downstream queries.} Each query processing unit can invoke the query API of other units.
  \item \textbf{Callback computation.} When a QPU invokes the query API of another, a stream connection is
  established between them, as described above.
  The unit implements a \textit{callback computation} that is invoked in response to receiving a record through that
  stream.
  \item \textbf{Configuration state.} Each QPU maintains additional configuration that specify its query
  processing capabilities.
  For example, in the case of an index QPU class this may include configuration parameters specifying which
  attribute it is responsible for indexing (and hence which queries it can process).
  In addition, it includes information that enables it to send downstream queries to other units, such as their
  endpoints and query processing capabilities.
\end{itemize}

\begin{figure}[t]
  \centering
    \includegraphics[width=0.3\textwidth]{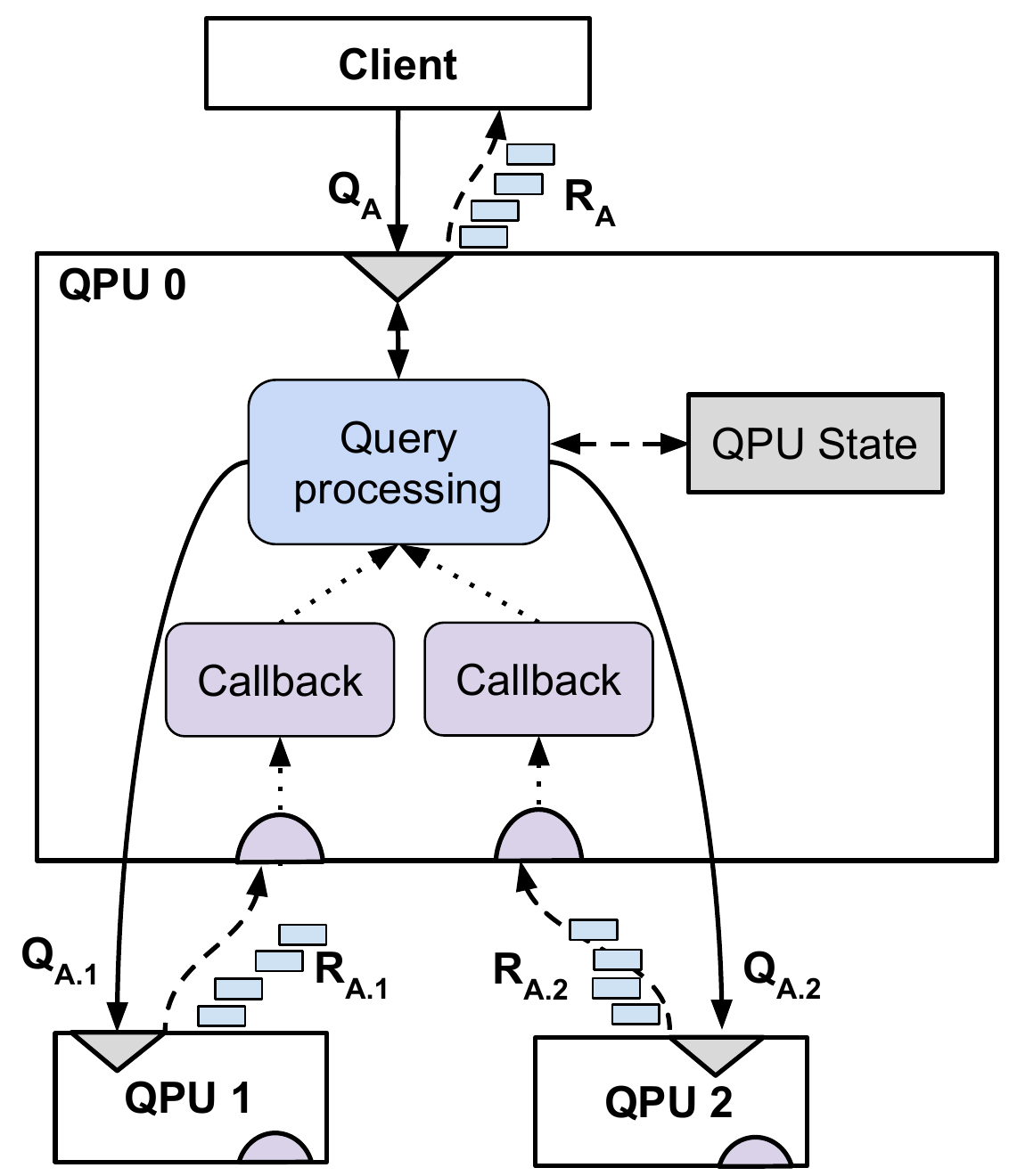}
  \caption{A conceptual depiction the QPU abstraction.}
  \label{fig:qpu}
\end{figure}

A conceptual depiction of the query processing unit abstraction is shown in Figure~\ref{fig:qpu}.
When the QPU's query API is called, a response stream ($R_A$) is established between the unit and the client, and
the unit's query processing computation is invoked.
The query processing computation can read the QPU's state, and can perform downstream queries to other units.
For each downstream query, a corresponding stream is established ($Q_{A.1}$ and $Q_{A.2}$).
When a record is received from one of the streams, the QPU's callback computation is invoked.
Each callback computation processes the received record, and returns the result to the query processing computation.
Upon receiving a result from the callback, the query processing computation can write to the QPU's
state and potentially send a computed query result through the response stream.

\subsection{Query processing system architecture}
A query processing system is a directed acyclic graph (DAG) with QPUs as its nodes.
Edges are connections between QPUs, which represent potential paths of communication among them:
the QPU at a parent node can perform a downstream query to the QPU at the child node.
Leaf nodes communicate with the data storage tier, while clients perform queries by invoking the query API of root
nodes.
The global capabilities of a QPU-based DAG are synthesized from the individual functionalities of
each of the QPU classes at its nodes, as well as the graph topology.

\subsection{Computation model}
The QPU graph runs a distributed bidirectional data-flow computation.

A client performs a query $Q_c$ by invoking the query API of a query processing unit at the root of the graph.
As described in Section~\ref{subsec:qpu}, the QPUs query processing computation can read from the unit's state,
or perform downstream queries to QPUs at its child nodes.

When a downstream query is performed, this process is recursively executed at each unit whose query API is invoked.
Though this mechanism, $Q_c$ is incrementally transformed to sub-queries which flow downwards through the QPU graph,
invoking computations at different nodes.
Sub-query results are returned through the QPU streams established from query API invocations, and flow upwards
through the graph.
These results are incrementally processed, potentially updating the state of different QPUs, and eventually
produce the initial query results, which are returned to the client.

\subsection{Constructing a QPU-based Query Processing System}
\label{sub_sec:qpu_usage}
In this section we describe the process of constructing and deploying a QPU-based query processing system, and
show how this design pattern achieves our objectives.

The process of constructing a query processing system consists of four steps.
The first step is selecting the QPU classes to be used for implementing the query processing functionalities
needed by the application.
The second step is designing the QPU DAG topology by defining the instances of QPU classes to be used as
nodes, and the connections among them.
The third step is defining the placement of each graph node across the system infrastructure.
The final step is deploying the QPU graph.
This step involves deploying each process that implements a QPU class instance, and passing to it configuration
that defines each functionality and its children in the QPU graph.

The properties of the query processing unit abstraction, and the properties that emerge from the composition of
QPUs realize our design objectives.

\textbf{Declaratively defined architecture.}
The properties of the QPU abstraction (common query API,
downstream queries) make query processing units composable.
This enables a query processing system to be expressed as a composition of building blocks (QPUs) that provide basic
query processing functionalities.
Different query system architectures can be expressed by selecting the QPU classes to be used and the topology of
the connections among them.

In addition, the common query API across any QPU class enables the separation between interface and
implementation.
For example, different implementations of an index QPU class can use different index data structures.
More generally, a graph node is agnostic of the sub-graph below each of its children and only requires local information
about the query capabilities of its children.
For example, a QPU class that implements query result caching can be transparently connected to the root of any
QPU sub-graph, such as an individual index QPU or a sub-graph that implements a partitioned index.

\textbf{Flexible component placement.}
Query processing units act as microservices:
each unit manages its internal state, and communication is performed through API invocations.
Therefore, each individual QPU of the query system DAG can be independently placed across the system, without
impacting the system's functionality.
This property decouples the architecture design from the placement of its components, and gives the system designer
control over the placement the query system's state and computations.


\section{Case study: Materialized Views At the Edge}
\label{sec:mv_edge}
\subsection{Motivation}
We consider an application that provides an in-game advertisement service for mobile game development.
Game applications use this service to query a database for available advertisement videos (ads), select an ad based on some
criteria, and then request the ad from the service in order to display it to the user.
We are interested in the part of the application's architecture responsible for serving queries to the ad database, and
consider the video serving architecture out of the scope of this study.

Each ad is associated with a set of tags describing its content, and a price.
The ad-serving application and game developers are paid per-click according to the ad price.
Tags are used to select ads with content that users are likely to click on.

Advertisers perform writes to the database to add or remove ads, and to adjust ad prices.
Because the available ads and their prices are modified frequently, ad selection is performed at
the point of the game when an ad needs to be displayed.
Therefore, queries require low latency as they impact user experience.
Additionally, queries require fresh results, as having the latest information about available ads and prices
enables ad selection to make optimal choices for maximizing profit.

\subsection{Background}
The ad-serving application's storage tier consists of a database that stores the available ads, their associated tags
(table Ads), and their prices (table Prices).

We are interested in the design of the application's query processing tier.
Our goal is to achieve low query response time while also providing fresh query results.
The common-case query performed by games has the following characteristics: (1) it selects ads that contain tags among
a given set, (2) joins the selected ads with the associated prices, and (3) selects the ads with the $K$ highest prices.

A technique for providing low query response time is to use materialized views \cite{Noria2018}.
However, existing approaches execute queries in the data center.
The ad-serving application's users are distributed worldwide, and therefore, the communication latency between user
devices and the data center may be significant.
Placing data geographically closer to end users is a common technique for reducing the large access latencies resulting
from geo-distribution \cite{google:infra}.

\subsection{Design decisions and trade-offs}
We assume a system composed of two types of \textit{sites}: data centers and edge sites.
The storage tier is placed in the data center, while the query processing tier can be distributed across data
center and edge.
For simplicity, for the rest of this section, we focus on a single data center and edge site.

The query system is queried by clients located at the edge;
it keeps its derived state up to date by asynchronously receiving updates from the database,
which is located in the data center.
As a result, query performance is affected by the latency between the query system's state and clients,
while the consistency between base data and the query system's state is affected by the latency between them.

There exists thus an inherent trade-off between query performance and freshness which is affected by the placement
of the query system's state relative to the clients and the base data.
Placement at the edge can improve query performance at the expense of query result freshness.
Therefore, it is more suitable for parts of the state that are heavily queried.
On the other hand, placement in the data center is better suited for parts of the state that are more frequently updated.

Additionally, placement decisions affect network consumption costs.
When the query system's state is placed in the data center, costly inter-site communication is required for sending
queries and responses between the query system and the clients.
When it is placed at the edge, inter-site communication is required for propagating updates from base data to
the query system.

\subsection{System design}
In this section, we present the design of a QPU-based query processing system for the ad-serving application
described above.
We demonstrate how the proposed architectural design pattern enables flexible placement of the query system's state
across data center and edge.

\textbf{QPU Classes.}
Following the steps defined in Section~\ref{sub_sec:qpu_usage}, we first define the QPU classes that will be
used for constructing the query system.
\begin{itemize}
  \item \textbf{Data store driver QPU (DSD-QPU).}
  The DSD-QPU is responsible for connecting the query processing system DAG with the underlying database.
  Its query API provides the following functionality.
  Given a query, the DSD-QPU initially returns the query result (by sending each result entry through the response
  stream).
  It then continues publishing (as stream records) changes to the database state that affect the result of the given
  query.
  \item \textbf{Index QPU (I-QPU).}
  The I-QPU is responsible for maintaining an index for a database table column.
  Upon deployment, the index QPU performs a downstream query to a DSD-QPU that is responsible for the
  corresponding column.
  It initially receives the entire column and builds its index.
  It then incrementally updates the index by receiving updates for database writes from the DSD-QPU.
  Its query processing computation processes a given query by performing an index lookup.
  \item \textbf{Join QPU (J-QPU).}
  The join QPU is responsible for performing the SQL join operation on two or more database tables or materialized
  views.
  Similarly to the index QPU, it performs downstream queries to other QPUs in order to build the join
  operation result, and keep up to date by incrementally applying updates.
  \item \textbf{Top-K QPU (TK-QPU).}
  The top-K QPU is responsible for maintaining a materialized view with the $K$ records of a database table or
  materialized view with the highest values in a given column.
  Similarly to the index and join QPUs, it uses downstream queries to implement this functionality.
  \item \textbf{Cache QPU (C-QPU).}
  The cache QPU is responsible for storing recent query results in a cache.
  When a cache miss occurs, it forwards the query further down the graph through a downstream query.
\end{itemize}

\begin{figure}[t]
\centering
\begin{minipage}{.5\columnwidth}
  \centering
  \includegraphics[scale=0.45]{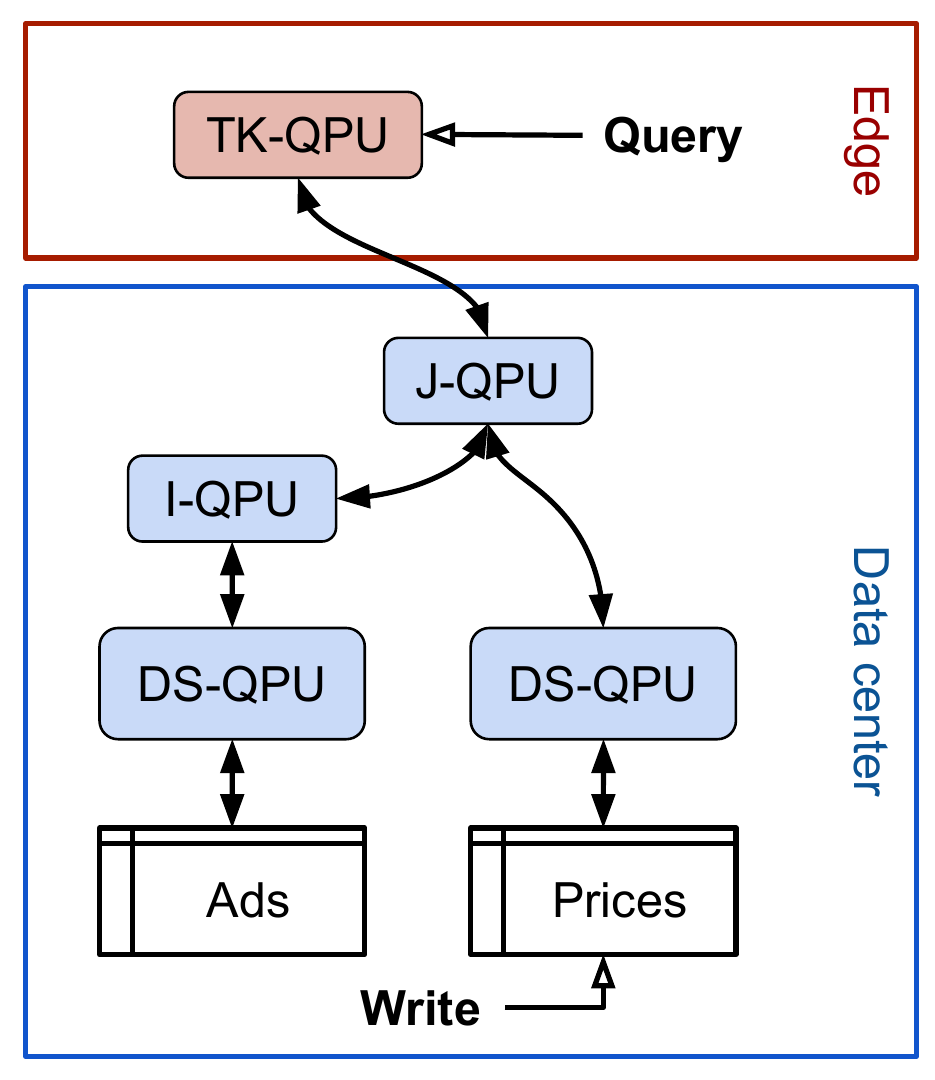}
  \subcaption{}
  \label{sub_fig:mv_edge_1}
\end{minipage}%
\begin{minipage}{.5\columnwidth}
  \centering
  \includegraphics[scale=0.45]{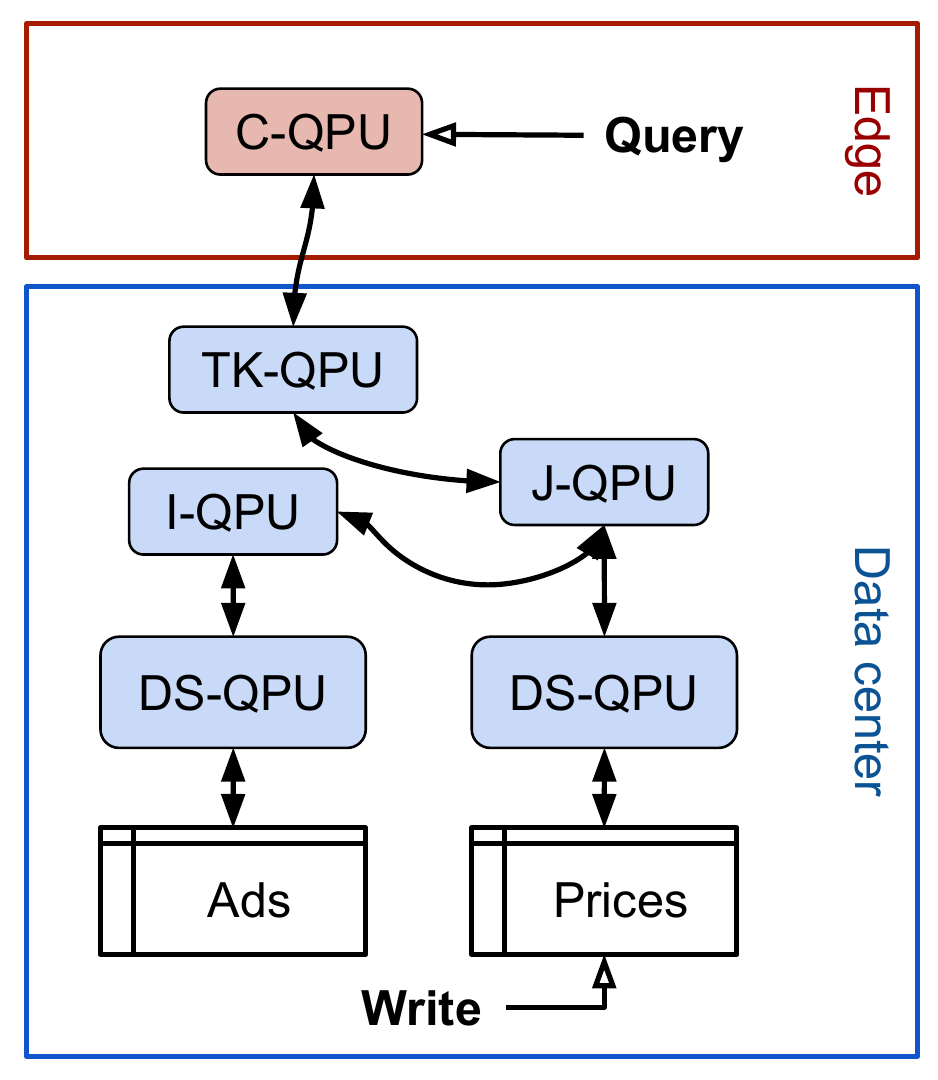}
  \subcaption{}
  \label{sub_fig:mv_edge_2}
\end{minipage}
\caption{Alternative query system designs and placement schemes for the in-game advertising application, addressing different workload types.}
\label{fig:mv_edge}
\end{figure}

Figure~\ref{fig:mv_edge} depicts the two query processing system architectures for the application.
The index QPU maintains an index for the ad tags.
The join QPU performs a join between the index and the $Prices$ table.
As a result of the join, index entries contain ads and corresponding prices.
The top-K QPU maintains, for each index entry, the $K$ ads with the highest prices.

We present two alternative query systems, designed to address different workloads.
In Figure~\ref{sub_fig:mv_edge_1}, the top-K QPU is placed at the edge.
This design is better suited to query-heavy workloads, as it favors query performance but sacrifices query result
freshness.
In Figure~\ref{sub_fig:mv_edge_2}, the top-K QPU is placed in the data center, and a cache QPU is placed at
the edge.
This design is more suitable for update-heavy workloads.
Placing the top-K QPU in the data center achieves better freshness, while the cache QPU placed at the edge
can potentially speed up query processing.


\section{Related Work}
\label{sec:related}
CockroachDB, a globally-distributed SQL database, provides different choices for distributing and placing data across
multiple data centers, termed topology-patterns, aimed at reducing read and write latency and improving resiliency \cite{CockroachDB:topology}.
However, these placement options are primarily focused on the data storage tier, rather that the query processing system.

The indexing systems in Azure DocumentDB \cite{10.14778/2824032.2824065} and Diff-index \cite{DBLP:conf/edbt/TanTTF14}
allow database administrators to choose between multiple index update modes.
While this enables a trade-off between index consistency and the overhead incurred to
update operations, these works do not consider querying processing across multiple data centers.

Noria \cite{Noria2018} is a middleware system aimed at improving performance of read-heavy web applications.
Similarly to our approach, Noria uses data-flow to incrementally apply writes to materialized views.
However, Noria does not consider geo-distributed data, and is not focused on flexible materialized view placement.



Our previous work \cite{10.1145/3194261.3194265} addressed the problem of building modular query processing systems and
introduced the term Query Processing Unit.
This paper advances this work by presenting a detailed specification of the QPU abstraction and demonstrating how
it can be used to enable flexibility in the design of query processing systems.


\section{Conclusions}
\label{sec:conc}
In the context of geo-distribution,
the placement of query processing state affects query response time, query result freshness, and operational cost.
Different design choices cater to different use cases due to applications' diverse requirements and characteristics.

To address this challenge, we proposed an architectural design pattern for constructing query processing systems in a
per use case basis.
To realize this, we introduced the Query Processing Unit abstraction and showed how it can be used as a building block to
enable the construction of application-specific query systems.

In this work-in-progress paper, we presented the QPU abstraction and the modular system design pattern that it enables.
We have implemented the proposed approach in the form of a framework that includes (1) a library of QPU
implementations and (2) deployment, orchestration, and self-configuration mechanisms aimed to facilitate the
construction of query processing systems.
Our prototype implementation is available at \href{https://github.com/dvasilas/proteus}{github.com/dvasilas/proteus}.
We are currently in the process of experimentally evaluating our prototype.

In addition, we aim to extend this work with mechanisms for reducing the complexity exposed to the application
developer.
In particular, we are developing a cost model based approach for generating and deploying query engine architectures
based on use case descriptions that include the topology of the underlying storage system, the expected workload
characteristics, and the application's requirements in the form of a target metric (query performance, freshness or cost)
to optimize for.


\bibliographystyle{abbrv}
\bibliography{refs}

\end{document}